\documentclass[a4paper,twocolumn,aps,superscriptaddress,floatfix,prmaterials]{revtex4-1}
\usepackage{graphicx}
\usepackage{comment}
\usepackage{amsmath,amsfonts,amssymb,amsthm}
\usepackage{mathtools,braket}
\usepackage{url}
\usepackage{siunitx}
\usepackage{hyperref}
\usepackage{xcolor}

\newcommand{\be}{\begin{equation}}
\newcommand{\ee}{\end{equation}}
\newcommand{\ba}{\begin{eqnarray}}
\newcommand{\ea}{\end{eqnarray}}
\newcommand{\nn}{\notag}

\newcommand{\lb}{\label}

\newcommand{\eps}{\epsilon}

\newcommand{\bk}{\mathbf{k}}
\newcommand{\bd}{\mathbf{d}}
\newcommand{\br}{\mathbf{r}}
\newcommand{\bE}{\mathbf{E}}
\newcommand{\dt}{\partial_t}
\newcommand{\kpump}{\mathbf{K}_{\mathrm{pump}}}

\newcommand{\op}[1]{\hat {#1}}
\newcommand{\commutator}[2]{\left[ {#1} , {#2} \right]}
\newcommand{\trace}[1]{\mathrm{tr}\left(#1\right)}

\newcommand{\dm}{\op{\rho}}

\begin{document}

\title{Real-time modelling of Optical orientation in GaAs: generation and decay of the degree of spin polarization}
\author{M. D'Alessandro}
\affiliation{Istituto di Struttura della Materia-CNR (ISM-CNR), Via del Fosso del Cavaliere 100, 00133 Roma, Italia}
\author{D. Sangalli}
\affiliation{Istituto di Struttura della Materia of the National Research
Council, Via Salaria Km 29.3, I-00016 Montelibretti, Italy}
\date{\today}

\begin{abstract}
We present a real-time abinitio description of optical orientation in bulk GaAs due to the coupling with an ultrashort circular polarized
laser source.
The injection of spin polarized electrons in the conduction band is correctly reproduced,
and a non vanishing spin polarization ($\mathbf{P}$) parallel to the direction of propagation of the laser ($z$) emerges.
A detailed analysis of the generation and the evolution of $\mathbf{P}(t)$ is discussed. 
The single $\bk$-point dynamics is a motion of precession around a fixed axis with constant $|\mathbf{P}|$ and fixed frequency.
Instead, the $\bk$-integrated signal shows only a time dependent $P_z(t)$ and decays few pico seconds after the end of the laser pump due to decoherence.
Decoherence emerges since the individual contributions activated by the pump give rise to destructive interference. 
We interpret the results in terms of the \emph{free induction decay} mechanism proposed some years ago~\cite{Wu2010}.
For the first time we are able to reproduce such effect in a full
abinitio fashion, giving a quantitative estimate of the associated decay time. Our result also shows a possible explanation to the time decay of spin magnetization observed in many real-time abinitio simulations.
\end{abstract}

\maketitle

\section{Introduction}

The control of the spin degree of freedom by optical means is the key
to the realization of ultra-fast magnetic devices, where the reading
and writing of information could be performed on the femto-second time
scale.
This is why, following the first experimental demonstration of ultra-fast optical demagnetization in
nickel~\cite{Beaurepaire1996}, a very reach activity started, both experimental and theoretical,
with the emerging of a new research area, called femto-magnetism~\cite{Kirilyuk2010}.  Different physical effects
are involved in the indirect interaction between spin and light, and in all of these spin-orbit coupling (SOC) plays a crucial role~\cite{Wu2010}.
While most of the studies focus on magnetic or anti--ferromagnetic materials, the control of the spin degree of freedom can be achieved in para--magnetic materials via circularly polarized light.
An example is optical orientation~\cite{Zutic2004,Averkiev2008,Meier2012optical} in GaAs, due to
the interplay between SOC and the space group of the crystal which determines the selection rules for light
absorption at the $\Gamma$ point of the Brilloiune zone (BZ).
If GaAs is excited with a circularly polarized laser pulse, whose frequency is tuned close to the optical gap,
an unbalance between the spin--up and the spin--down
electronic density in the conduction band is generated. Such effect was known~\cite{Pierce1976} long before the realization of the first ultra-fast experiment on the femto-second timescale and it was studied in particular in GaAs quantum wells~\cite{Pfalz2005}.
It already has technological
interest by its own, for the generation of spin polarized currents for spintronics devices.
The underlying physics can be captured
by a simple 6 states model~\cite{Pierce1976,optorient1984,Nastos2008,Rioux2012} (8 states are needed if the split-off bands need to be included). The degree of
polarization predicted by the model in the conduction band is 50\%, in good agreement with experimental results~\cite{Meier2012optical}.
The few states model has
the advantage of clearly capturing the physics of optical orientation, giving a simple interpretation
in terms of
populations in the different bands of GaAs. A description beyond such simple model can be achieved via the
$\mathbf{k\cdot p}$ approach, which also describes how the degree of spin polarization changes moving away
from the $\Gamma$ point, i.e. tuning the laser frequency above the optical gap value.
A fully abinitio description was formulated more recently~\cite{Nastos2007}, studying the second order
response derived from the equation of motion for the density matrix. The approach of Ref.~\onlinecite{Nastos2007}
highlighted the role played by coherences between SOC spin-split bands in the conduction.
The SOC spin-splitting is due to the lack of inversion symmetry in GaAs and exists
only if bands dispersion away from the high symmetry lines is considered.
Coherences between such spin-split bands
account for $\approx 70\%$ of the spin polarization induced by the optical pulse.

While these works clarified the physics behind optical orientation, they remain confined to the description in
frequency space. In view of the renewed interest in spin and spin currents for femto-magnetism applications~\cite{Zhang2013,Tancogne-Dejean2020}, an
approach in the time domain is instead needed.
In this work we adopt a fully abinitio real time approach, and
we show that it captures the correct value of the degree of spin polarization in the conduction band (and also in the
valence band). We are able to follow its coherent generation and predict that the super-imposed dynamics of
the coherences are responsible for a subsequent decay on a time scale of few ps.

In the time domain the key question is indeed how the spin polarization evolves and decays~\cite{Kikkawa1997,Kikkawa1998}.
The abinitio realization of real time spin dynamics is a new and very exciting approach, which has seen renewed interest in recent years~\cite{Krieger2015,Shokeen2017,Simoni2017}, for the possibility to shine new light on such key questions.
However, the connection between these recent works and the broad and well established literature of spin relaxation models and mechanisms~\cite{Shah1996,Gardiner2004,Wu2000_epjb,Wu2000_PRB,Weng2003,Stich2007} is not always clear. The two most established
spin decay channels~\cite{Wu2010} are the Dyakonov-Perel' (DP) and the Elliot-Yaffet (EY) mechanisms, which are both related
to the interplay between SOC and scattering. The dynamics due to the DP is dictated by the ratio between the spin precession time $1/\Omega$, i.e. the characteristic time frequencies of the coherences in our language, and the scattering time $\tau_p$.
In the weak scattering regime, $\Omega \tau_p >1$, spin precession leads to \emph{free induction decay} in presence of a continuous distribution of frequencies (\emph{inhomogeneous broadening})~\cite{Wu2010,Slichter1990}.
We will interpret our results taking advantage of these concepts to put a bridge
between the abinitio community and the physics of models.

Let as remark, as already discussed in the literature~\cite{Nastos2007},
that a critical aspect in the description of optical properties in
GaAs is the k-sampling of the BZ.
In this work we adopt a specific ultra-fine sampling of random points around the $\Gamma$ point. The size of the region is chosen according
to the frequency profile of the laser pulse used in the real-time propagation.

\section{Optical orientation}

The time dependent magnetization of optically excited electrons in the conduction band can
be probed in time-resolved photo-emission (TR-PE) experiments.
In the present case GaAs is driven out of equilibrium by a femto-second circularly polarized laser pulse, the pump, whose frequency is tuned at
\SI{1.5}{\eV}, slightly above the direct gap of GaAs. To fix the
geometry we define $z$ as the direction of propagation of the field,
which is then circularly polarized in the $x$-$y$ plane. We consider a
pulse whose envelope is a Gaussian function with a FWHM of
\SI{100}{\fs}. This corresponds to an energy spread of the pulse of
around \SI{40}{\meV}.

Since spin is a one body operator, the dynamics of the ``degree of spin polarization'' for electrons excited in the conduction band, $P^c_z(t)$ can be expressed in terms of the time-dependent one-body reduced density-matrix $\dm(t)$ as follows:
\be
P^c_z(t) = \frac{\trace{\dm^{c}(t)\op{S}^z}}{\trace{\dm^c(t)}} \, .
\label{eq:Pcz_def}
\ee
Here $\op{S}_z$ is the $z$ component of the spin operator, ${\dm^c(t)=\op{\Pi}_c\dm(t)\op{\Pi}_c}$ is the projection of the non-equilibrium density matrix in the conduction subspace via the projector $\op{\Pi}_c$.
Eq.~\eqref{eq:Pcz_def} can be obtained from the definition of the PE
signal, assuming a probe pulse longer than the very fast oscillations
of $\dm$ in the $\{cv\}$ channel, and with enough energy to extract
electrons only from the conduction band.
We observe that $P^c_z$ defined according to Eq.~\eqref{eq:Pcz_def} is
an intensive quantity that measures the ratio between the expectation
value of the spin operator, restricted in the conduction subspace, 
and the number of carriers.

The time-dependence of Eq.~\eqref{eq:Pcz_def} is codified in the equation of motion (EOM) for $\dm(t)$ which can be derived from the Kadanoff-Baym equation
under the Generalized Kadanoff--Baym approximation. We express $\dm(t)$ in the basis of the Kohn-Sham (KS) states
\be\lb{dm_ksbasis}
\dm = \sum_{nm\bk}\rho_{nm\bk}\ket{n\bk}\bra{m\bk} \, ,
\ee
where the band indices $n,m$ run over the KS states, both occupied and
empty, and $\bk$ is the Bloch momentum. The EOM for matrix elements of 
$\dm(t)$ in the KS basis can be expressed as
\be\lb{eq:dm_eom1}
i\dt\rho_l = \eps_l\rho_l +\commutator{\Delta\op\Sigma[\dm]}{\dm}_l +
\commutator{\op{U}^{\mathrm{pump}}}{\dm}_l + S_l[\dm] \, ,
\ee
where $l = \{nm\bk\}$ is a compact multi--index notation and
$\commutator{\op{O}}{\dm}_l$ is the matrix element of the commutator
between the operator $\op{O}$ and the DM expressed in the KS basis.
$\op{U}^{\mathrm{pump}}$ describes the pump pulse, \emph{i.e.} the
source term that drives the system out of equilibrium.
$\Delta\op{\Sigma}[\dm]$ is the self-consistent variation of the
static part of the self-energy, including the mean field, while
$\op{S}[\dm]$ captures the variation in the dynamical part of
the self-energy, including possible dissipation and relaxation
effects.

In the present analysis we work in the independent particle approximation, thus $\Delta\op{\Sigma}[\dm]=0$.
We also neglect all dynamical many--body effects and simplify the scattering term to a damping in the in the
$\{cv\}$ channel: $S_{nm\bk}[\dm]=-\eta_{cv}|f^{eq}_{n\bk}-f^{eq}_{m\bk}|\rho_{nm\bk}$.
This term is needed to \emph{cure} the finite nature of the BZ sampling
and to provide converged results (further details on its choice are
discussed in the next sections). Instead, no decoherence term is imposed in the $\{cc\}$ channel and, accordingly, $P^c_z$ defined in
eq.~\eqref{eq:Pcz_def} fully contains the coherent signal.

\begin{figure}[t]
\begin{center}
\includegraphics[scale=0.32]{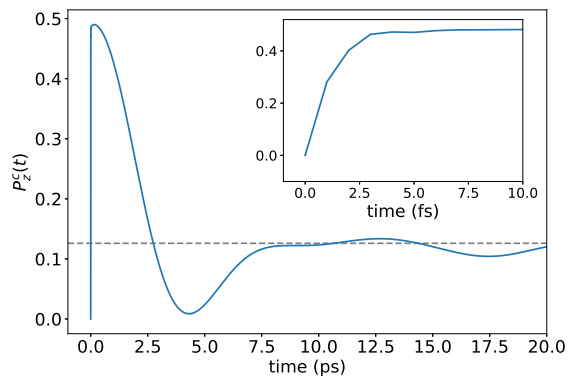}
\end{center}
\caption{Degree of spin polarization as a function of time. The horizontal dashed line represents the non coherent contribution due to the populations only, \emph{i.e.} considering only the diagonal elements of the density matrix.}
\label{fig:Pz_result}
\end{figure}

Equation \eqref{eq:dm_eom1} has been solved in the time domain by using a
development version of the YAMBO package~\cite{Marini2009,Sangalli_2019},
that evaluates the temporal evolution of $\rho_{l}(t)$ through a $2^{nd}$
order Runge-Kutta integrator. 
The resulting $P^c_z$ computed propagating eq.~\eqref{eq:dm_eom1} on a fine random grid centered around the $\Gamma$ point is shown in Fig.~\ref{fig:Pz_result}. Details of the simulations are discussed in the next sections.
A value for $P_z^c$ very close to the theoretical result $P_z^{c,theo}=0.5$ is
reproduced as soon as the laser is switched on (it takes just few fs, i.e. about a optical cycle of the laser, see inset).
The little discrepancy with respect to the results of the simple
model~\cite{Pierce1976} at $\Gamma$ is easily understood since the real-time
pump, due to its energy and finite time width, activates also transitions 
aside from $\Gamma$, where the ratio between spin up and spin down 
transitions is not exactly 3:1.
$P_z^{c}(t)$ however drops, on the time--scale of few ps, despite no
decoherence mechanism is included in the EOM for the ${cc}$ channel. 
Indeed, the coherent contribution drops to zero due to the \emph{free 
induction decay}, despite all $\rho_{cc'\bk}(t)$ terms remain finite and
oscillates in time at $\Delta\epsilon_{cc'\bk}$. We refer to the residual 
spin polarization, $P_z^{c}(\infty)$, as a ``dephased limit'' . 
$P_z^{c}(\infty)\approx P_z^{c,diag}$, the value obtained imposing 
$\rho_{cc'\bk}=0$ if ${c\neq c'}$.
$P_z^{c,diag}\approx 0.125$ is less than 30\% of $P_z^{c,theo}$ as predicted
in the literature~\cite{Nastos2007}, and is marked by the horizontal dashed
line in the plot.

\subsection{Preliminary analysis of the band structure}
\lb{sec:bz_sampling}

\begin{figure}[t]
\begin{center}
\includegraphics[scale=0.32]{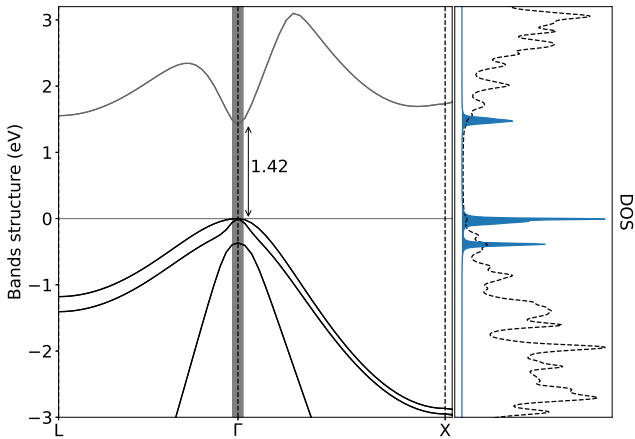}
\end{center}
\caption{(Left) Band structure including the scissor correction along the $L-\Gamma-X$ path. The shaded gray slice close to $\Gamma$ denote the $\kpump$ optical active region. (Right) The black (dashed) curve represents the DOS of the system. The blue (shaded) are shows the DOS restricted to the $\bk$ points in the
optical active region. The two curves are rescaled for better visualization.}
\label{fig:bands_LGX}
\end{figure}

KS energies and wave functions are computed,
with a non collinear approach and
using the PBE~\cite{Pbe1996} exchange and correlation functional,
as implemented in the QuantumESPRESSO package~\cite{Giannozzi_2009,Giannozzi_2017}.
SOC is included via fully relativistic pseudo-potentials for both Ga and As.
The ground state of the system is obtained with the following
convergence parameters: a $6 \times 6 \times 6$ Monkhorst-Pack $\bk$ grid and an energy cutoff of \SI{80}{Ry}, that ensure convergence in
the total energy at the level of a fraction of \SI{1}{\meV}. 
The equilibrium lattice constant for the given description of the GaAs
lattice has been computed by relaxing the system. The lattice parameter
that provides a stable configuration is
$a_{lat}=\SI{5.54}{\angstrom}$.
Empty bands are evaluated performing non self-consistent computations
based on the converged density.
As expected, PBE underestimate the (direct) band gap.
We use a scissor correction of \SI{0.546}{\eV} to match
the ``experimental'' band gap of \SI{1.42}{\eV} at
\SI{300}{\kelvin}.
The corrected band structure along the $L-\Gamma-X$ high symmetry path is shown in Fig.~\ref{fig:bands_LGX}. Along such path each band is twofold degenerate
and can accommodate both ``spin-up'' and ``spin-down'' electrons.
In the valence sector we recognize the heavy and light holes, together
with the split-off bands lifted down by the SOC.

\begin{figure}[t]
\begin{center}
\includegraphics[scale=0.35]{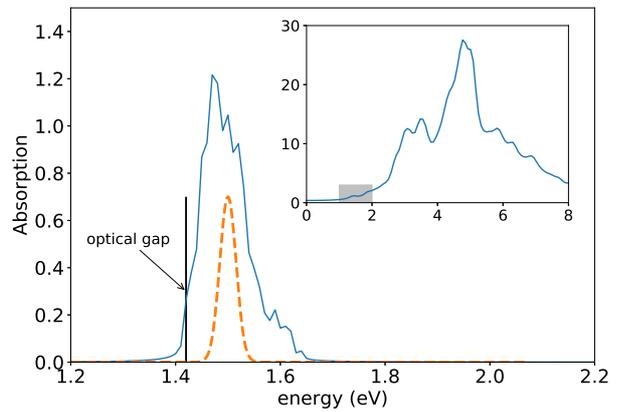}
\end{center}
\caption{(Blue line) Absorption from the top of the valence bands due to transitions from the heavy/light holes to the lower conduction bands.
The orange (dashed line) describes the energy profile of the pump, centred at
\SI{1.5}{\eV}.
The inset shows the absorption associated to all the allowed transitions in a wider energy range. The grey
shaded area of the insets is the part of the spectrum represented in the main frame.}
\label{fig:abs_cv}
\end{figure}

In view of the solution of Eq.~\eqref{eq:dm_eom1} we need a fine sampling which includes the \emph{optical active region} of the BZ,
i.e. all the transitions $\Delta\epsilon_{cv\bk}\approx\omega_0\pm\Delta_0$
that can be activated by the pump.
Given the frequency profile of the pump pulse, 
we chose a cube, denoted
as $\kpump$, with edge of 0.05 (in units of $2\pi/a_{lat}$) centred at $\Gamma$.
$\kpump$ is highlighted in Fig.~\ref{fig:bands_LGX} with a grey shadow, and 
it is sampled with $N_k=24000$ points, generated expanding by symmetry an initial set of $500$ random points with a uniform distribution.
The symmetry expansion is needed since only specific portion of the BZ are activated by the pump pulse, and to avoid spurious symmetry breaking in the simulations.
All the 24000 points lie within $\kpump$ and $\rho_{nm\bk}(t)$ is defined and propagated for each of them. If the whole BZ had been sampled
with the same density \num{4.8e7} $\bk$ points would have been needed.
To understand why such a fine sampling is needed,
the absorption, $\Im[\epsilon(\omega)]$, from the light and heavy holes states to the first conduction within $\kpump$, is shown in Fig.~\ref{fig:abs_cv},
and compared with the frequency profile of the laser pulse.
$\Im[\epsilon(\omega)]$ can be obtained from the Fourier transform of the polarization, $\mathbf{P}(t)=\sum_{cv\bk} \mathbf{x}_{cv\bk}\rho_{cv\bk}(t)$.
In order to overcome unphysical delta-shaped structures due the finite nature
of the sampling, and to ensure that the pump pulse is correctly absorbed,
the $\rho_{cv\bk}(t)$ matrix elements are dephased by a parameter $\eta_{cv}$,
which transforms each peak into a Lorentzian function.
Due to the nominal width of the laser pulse of $\SI{40}{\meV}$, we set 
$\eta_{cv}=\SI{4}{\meV}$. The reason of this choice is a
balance between faster convergence (bigger $\eta$) and letting the spread of optically activated $\bk$--region to be determined by the laser pulse. 
Fig.~\ref{fig:abs_cv} highlights that the chosen sampling and the size of 
$\kpump$ are adequate. In the figure we also show that the pump is tuned 
$\SI{80}{\meV}$ above the optical gap.

\begin{figure}[t]
\begin{center}
\includegraphics[scale=0.35]{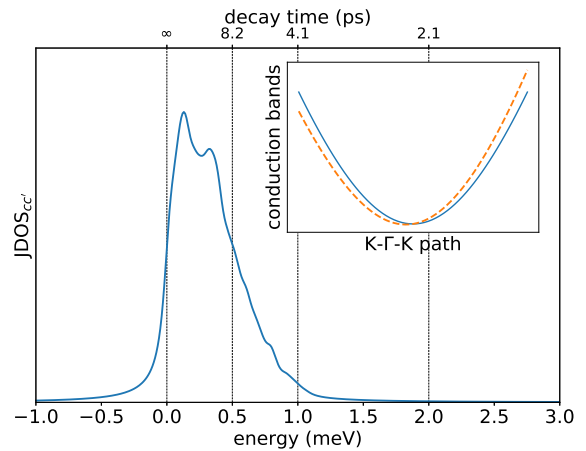}
\end{center}
\caption{JDOS$_{cc'}$ of the direct transitions between the spin-splitted states of the conduction bands. 
The insets contains a representation of the conduction bands along the $K-\Gamma-K$ path close to the $\Gamma$ point.}
\label{fig:jdos_ccp}
\end{figure}

Although the first two conduction bands are degenerate along the
$L-\Gamma-X$ path (see Fig.~\ref{fig:bands_LGX}), this is not the case for other $\bk$ points in the $\kpump$ region. A detailed analysis~\cite{Gmitra2016}
shows that the SOC, together with the lack of inversion symmetry, is responsible for a spin splitting away from the high symmetry lines,
with $\epsilon_{\uparrow\bk}=\epsilon_{\downarrow-\bk}$ (see the inset of Fig.~\ref{fig:jdos_ccp}).
Let us now focus on such transitions in the conduction sector that will determine the dynamics of $\rho_{cc'\bk}(t)$.
The Joint Density of States in this sector (JDOS$_{cc'}$) is
reported in Fig.~\ref{fig:jdos_ccp}.
The energy differences ${\Delta\epsilon_{cc'\bk}\approx(\epsilon_{\uparrow\bk}
-\epsilon_{\downarrow\bk})}$ in the $\kpump$ region are of the order of few tenths of a meV and limited to a maximum energy of \SI{1}{\meV}.
We thus expect a dynamics on the ps time--scale associated to such energy differences
($\SI{1}{\meV}\rightarrow\SI{4}{\ps}$, see upper x-axis in Fig.~\ref{fig:jdos_ccp}).
Also in this case we need to choose a spread parameter, $\eta_{cc}$, in this channel
to balance between faster convergence
(big $\eta_{cc}$ and smoother JDOS$_{cc'}$), and avoiding to alter the dynamics
on the ps time--scale 
We use the value ${\eta_{cc}=\SI{0.05}{\meV}(\rightarrow\SI{80}{ps}})$ which is about a factor 10 smaller of what would impact the results of our real--time simulations but big enough to
have a smooth JDOS$_{cc'}$.

\subsection{Real time simulations}

The degree of spin polarization, Eq.~\eqref{eq:Pcz_def},
$P_z^c$ can be expressed explicitly in the KS basis, using
${\op{\Pi}_c = \sum_{c\bk}\ket{c\bk}\bra{c\bk}}$:
\be\lb{eq:mz_cond2}
P^c_z(t) = \frac{\sum_{cc'\bk}\rho_{cc'\bk}(t)S^z_{c'c\bk}}{\sum_{c\bk}\rho_{cc\bk}(t)} \, ,
\ee
where
\be\notag
S^z_{cc'\bk} = \frac{1}{2}\int d\br\left(
\psi^*_{c\bk\uparrow}(\br)\psi_{c'\bk\uparrow}(\br)-\psi_{c\bk\downarrow}^*(\br)\psi_{c'\bk\downarrow}(\br) 
\right) \, .
\ee
By considering the coupling with the pump of the form of a dipole
interaction, \emph{i.e.}
$U^{\mathrm{pump}}(t)=\mathbf{E}(t)\cdot\mathbf{r}$, we can recast
Eq.~\eqref{eq:dm_eom1} as a damped oscillatory equation driven by
a self consistent source
\be\lb{eq:dm_eom2}
\dt\rho_l +i\Omega_l \rho_l  = F_l[\rho] \, ,
\ee
where $\Omega_l = \eps_l -i\eta_l$ and $\eta_l$ is the dephasing parameter
presented in sec.~\ref{sec:bz_sampling}: $\eta_l=\eta_{cv}=\SI{4}{\meV}$ in 
the $\{cv\}$ channel and $\eta_l=\eta_{cc}=\SI{0.05}{\meV}$ in the $\{cc'\}$
one. The source term of Eq.~\eqref{eq:dm_eom2} reads
\be\lb{eq:dm_eom3}
F_{nm\bk}[\rho] = -i\bE(t)\cdot
\sum_{p}\left(\bd_{np\bk}\rho_{pm\bk}-\rho_{np\bk}\bd_{pm\bk}\right)
\, ,
\ee
where the $\bd_l$ are the matrix elements of the dipole operator in the KS basis.
Expanding the DM in powers of the pump, up to the two-photon term, we obtain
\be
\rho_{nm\bk}(t) = \delta_{nm}f^{eq}_{n\bk} + \rho_{nm\bk}^{(1)}(t)+\rho_{nm\bk}^{(2)}(t) +\cdots \nn \, .
\ee
For times preceding the activation of the pump we have set ${\rho_{nm\bk}=\delta_{nm}f^{eq}_{n\bk}}$, so that the DM is
defined by the equilibrium occupations of states. Plugging this expansion into equations
\eqref{eq:dm_eom2} and \eqref{eq:dm_eom3} provides a chain of differential equations
for the $\rho^{(i)}_l$:
\be\lb{eq:dm_eom4}
\dt\rho^{(i)}_{nm\bk} +i\Omega_{nm\bk} \rho^{(i)}_{nm\bk}  =
F^{(i)}_{nm\bk} \, ,
\ee
which can be hierarchically solved starting from the lowest order. At the one-photon
level the source term reads
\be\notag
F^{(0)}_{nm\bk} = -i(f^{eq}_{n\bk}-f^{eq}_{m\bk})\bE(t)\cdot \bd_{nm\bk} \, ,
\ee
and is not vanishing only for transitions between valence and conduction states.
The solution of equation \eqref{eq:dm_eom4} at one-photon level shows that only
the matrix elements $\rho^{(1)}_{cv\bk}$ with 
$\Delta\eps_{cv\bk} \sim \omega_{pump}$ are activated by the pump through a
resonance mechanism. After an initial transient regime, these terms exhibit
an oscillatory behavior suppressed by an exponential factor, \emph{i.e.} $
\exp(-i(\Delta\eps_{cv\bk}-i\eta_{cv\bk})t)$.

Going ahead in the hierarchy of equations \eqref{eq:dm_eom4}, the ${\rho^{(2)}_{nm\bk}}$ terms are activated by two-photon processes. Here, we are interested in the generation of ${\rho^{(2)}_{cc'\bk}}$, which are leading terms in the $cc'$ channel. The source term associated to transitions among conduction states reads
\be\lb{eq:dm2photon_source}
F^{(2)}_{cc'\bk} =
-i\bE(t)\cdot\sum_{v}\left[\mathbf{d}_{cv\bk}\rho_{vc'\bk}^{(1)}-\rho_{cv\bk}^{(1)}\mathbf{d}_{vc'\bk}\right] \, . 
\ee
The $\rho^{(2)}_{cc'\bk}$ terms are activated if any of $\Delta\eps_{cv\bk}$
or $\Delta\eps_{c'v\bk}$ lay within $\omega_0-\Delta_0$ and $\omega_0+\Delta_0$.
The process becomes even more effective if the condition is satisfied by both.
Then $F^{(2)}_{cc'\bk}$ contains \emph{fast} oscillating terms
modulated by an envelope with frequency 
$\Delta\eps_{cv\bk}-\Delta\eps_{c'v\bk}$ which couples resonantly with the transitions
$\Delta\eps_{cc'\bk}$, activating a \emph{slow} oscillatory response
of the system on the ps time scale.
We now inspect the terms for which $S^z_{cc'\bk}$ is maximum.
For $c=c'$, $S^z_{cc\bk}$ this happens when $\hat{S}^z \psi_{c\bk} \approx \pm 1/2 \psi_{c\bk}$, i.e. for eigenstates of $\hat{S}^z$.
Instead for $c\neq c'$, $S^z_{cc'\bk}$ is maximum for $\hat{S}^z \psi_{c\bk} \propto \psi_{c'\bk}$. Since $\psi_{c\bk}$ and $\psi_{c'\bk}$ have opposite spin directions, this happens for eigenstates of $\hat{S}^x$ or $\hat{S}^y$.
Notice that at $\Gamma$, and more in general whenever ${\Delta\epsilon_{cc'\bk}=0}$, one can choose an arbitrary rotation in the degenerate space so that only the diagonal terms survive.
Then everything is described in terms of eigenstates of $\hat{S}^z$, as in the standard interpretation of the 6 states model~\cite{Pierce1976,optorient1984}.
When ${\Delta\epsilon_{cc'\bk}>0}$ this is not possible anymore, and other states (including eigenstates of $\hat{S}^x$ or $\hat{S}^y$) must be involved.

\begin{figure}[t]
\begin{center}
\includegraphics[scale=0.35]{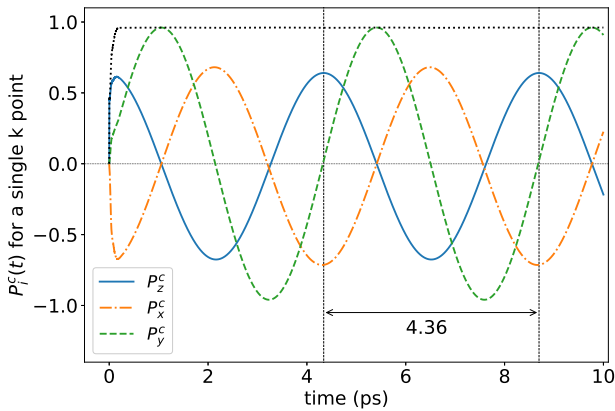}
\end{center}
\caption{Time profile of the components of the degree of spin polarization
associated to a single $\bk$ point, denoted $\bk_M$ in the text, with
cartesian coordinates $[\num{-0.025},\num{0.0076},\num{0.024}]$ in units of
$2\pi/a_{lat}$. The orange (dotted), green (dashed) and blue (continuous),
lines show the time behavior of the $x$, $y$ and $z$ components, respectively.
The black dotted line represents the time profile of the modulus of
$\mathbf{P}^c(t)$. All the components $P_i^c(t)$ oscillate with the same
frequency of \SI{4.36}{\ps}, reported in the figure.} 
\label{fig:spin_polarization_1k}
\end{figure}
In order to provide numerical support to these analytical considerations 
we have performed a real-time calculation including a single $\bk$ point.
For this test we have selected the point $\bk_M$ for which the transition
energy between $\Delta\eps_{cc'\bk} = \SI{0.948}{\meV}$ is the maximum,
among the points sampled in the $\kpump$ region. 
In Fig.~\ref{fig:spin_polarization_1k} we plot the contribution of such
point to $P^c_i(t)$ with $i=\{x,y,z\}$.
After a transient regime of the temporal range of the cycle of the
pump, the three components $P^c_i(t)$ oscillate with a frequency that exactly
matches the energy difference of the spin-split conduction bands 
($\SI{0.948}{\meV} \rightarrow \SI{4.36}{\ps}$).
The black dotted line shows that
$P^c=\sqrt{\sum_i (P^c_i)^2}$ remains constant during the time evolution. 
Moreover, we have verified that, if the contribution of all the points connected to $\bk_M$ by the symmetry are considered, then $P^c_x=P^c_y=0$ while $P^c_z(t)$ keep the same oscillatory behavior reported in the blue curve of Fig.~\ref{fig:spin_polarization_1k}.
Thus, while the single $\bk$ dynamics is captured by a precession around a fixed axis,
including symmetries the global dynamics results in a time oscillation in the modulus of
$\mathbf{P}^c$. This can be  visualized in the simplified case with two $\bk$ points and
\begin{eqnarray*}
\mathbf{P}^c(\bk_1)&=& P^c\left(0,+cos(\omega t), sin(\omega t)\right), \\
\mathbf{P}^c(\bk_2)&=& P^c\left(0,-cos(\omega t), sin(\omega t)\right).
\end{eqnarray*}

\subsection{Superposition of all k-points: emergence of a dephasing time}

We now turn to the global dynamics which result from the sum over 
the dense sampling of the single $\bk$ dynamics discussed in the previous
section. 
As shown in Fig.~\ref{fig:Pz_result}, the expectation value of $P^c_z(t)$
decays in the time-scale of few ps after the end of the pump. This behavior
emerges despite the fact that no dephasing mechanisms are present at this
time-scale and is due to the destructive interference among the elements of 
the spin ensemble in the $\kpump$ region which are actually activated by the
pump pulse.

We further inspect this mechanism through a numerical test. 
$P^c_z$ is the integral over $\bk$ of terms which oscillate as $cos(\omega_{\bk}t)$,
as shown in Fig.~\ref{fig:spin_polarization_1k}.
In Eq.~\eqref{eq:mz_cond2} the integral is performed as a simple sum.
Here instead, neglecting that each contribution has a different amplitude,
we carefully perform $\int d\omega\, cos(\omega t) j(\omega)$, where $j(\omega)$ is the JDOS$_{cc'}$ of Fig.~\ref{fig:jdos_ccp}.
To this end we use a trapezium method on a dense mesh of frequencies
interpolated from the initial 500 $\Delta\epsilon_{cc'\bk}$ abinitio energies.
\begin{figure}[t]
\begin{center}
\includegraphics[scale=0.35]{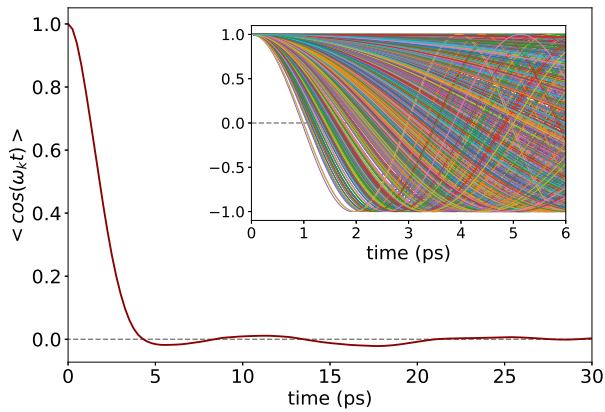}
\end{center}
\caption{Time profile of the average of oscillating functions with frequencies equal to the difference of the spin-split conduction bands.
The inset shows the profiles of the various addends.}
\label{fig:average_oscillations}
\end{figure}
The result is shown in the main frame of Fig.~\ref{fig:average_oscillations}. It has
a time profile very similar to the one of the $P_z^c(t)$ in Fig.~\ref{fig:Pz_result},
with complete decoherence taking place in  $\approx\SI{4}{\ps}$,
as a consequence of the Riemann-Lebesgue Lemma.

Lastly, we observe that this numerical test allows us to identify the 
residual long time oscillations of Fig.~\ref{fig:Pz_result} is due to the
finite number of $\bk$ points used to build $P^c_z$ in the real-time
approach. Indeed, for finite ensemble a finite portion of spin can be 
partially in phase after some time (revival of Rabi oscillations), whereas
in the continuum limit the amplitude of the long-time oscillations vanishes
since the revival time goes to infinity. 

\section{Discussion \& Conclusions}

We have proven that the dynamics of the degree of spin polarization, under 
the action of an ultra short laser pulse, can be fully captured within an
abinito approach.
The analysis of the single $\bk$-point dynamics shows that a spin precession
mechanism is activated. Such precession is determined by the Dressellaus field
connected to the magnitude and the direction of the SOC at such $\bk$,
as in standard Block models for spin dynamics~\cite{Wu2010}.
The spin polarization oscillates between $S_z$, $S_x$ and $S_y$.
In the superimposed dynamics of the symmetry connected $\bk$-points, which takes
into account the symmetries of the lattice external potential,
the $x$ and $y$ components of the spin polarization sum to zero at any time.
Thus the spin precession turns into a periodic transfer of angular momentum
from kinetic to potential rotational energy.
Only breaking such symmetries, for example via the use of an external magnetic
field, the spin-precession would appear in the macroscopic analysis.
The understanding of this mechanism gives a key instrument to interpret
recent works on spin-dynamics based on TDDFT in magnetic materials~\cite{Krieger2015,Shokeen2017,Simoni2017},where, similarly, only the $z$ component of the magnetization evolves in time.

The overall dynamics, due to the collective
response of the system, experiences a decoherence process on the time-scale of few $ps$,
due to the superimposed dynamics of periodic functions with multiple frequencies $\Delta\epsilon_{cc'\bk}$.
In the literature such process is named \emph{free induction decay} mechanism~\cite{Wu2010}.
Following standard notation, we refer to the global dephasing times
as longitudinal ($T^*_1$) and transverse ($T^*_2$).
In absence of external magnetic fields $T^*_1=T^*_2$.
Moreover we have $1/T^*_2=1/T_2+1/T'_2$, where $T_2$ accounts for explicit dephasing processes, which are negligible in our simulations, while $T'_2$ for \emph{free induction decay} mechanism, which we measure in the present work.
This is the first time, to our knowledge, that $T'_2$ is computed within a fully abinitio approach.
Given the chosen parameters for the pump pulse,
$T'_2$ depends on the JDOS$_{cc'}$ of GaAs, $\Delta\epsilon_{cc'\bk}$, in the subset of $\bk$-points activated by the pump.
Such subset contains all points for which $\Delta\epsilon_{cv\bk}\approx\omega_0\pm\Delta_0$.
We used a pump pulse detuned by \SI{80}{meV} from the optical gap 
of GaAs, $(\omega_0-\Delta\epsilon_{cv\mathbf{0}})\approx \SI{0.1}{eV}$ 
as shown in Fig.~\ref{fig:abs_cv},
and found that such detuning, rather than the value of $\Delta_0$,
determines the decoherence time under such conditions.
Indeed, even in the limit $\Delta_0\rightarrow 0$ of long laser pulses,
a single value $\Delta\epsilon_{cv\bk}$ is selected, which however corresponds to multiple
$\bk$-points and multiple values of $\Delta\epsilon_{cc'\bk}$.
We expect longer dephasing times for smaller detuning. 
A prediction which could be easily verified experimentally.

The present results also open the way to consider a number of extensions, like a detailed
analysis of the relation between the $T'_2$ and $\{\omega_0,\Delta_0\}$ for a wide range of
materials, or the study of the effects beyond the independent particles approximation. 
Let us conclude discussing how our results address one of the key question in ultra-fast
demagnetization experiments:
``\emph{where is the magnetization gone?}''.
In our case \emph{the magnetization is still there}.
The angular momentum is oscillating back and forth, with different frequencies.
However it is distributed over an infinite number
of degrees of freedom and it is \emph{not measurable anymore}.
Letting the atoms free to move, we would expect a periodic transfer of angular
momentum between the electrons spin $S_z$ to the atoms (or coherent phonons states).
As for the electronic case, the coherent phonons would likely experience
decoherence and no macroscopic momentum could be detected experimentally beyond few ps.

\subsection*{Acknowledgments}

We acknowledge funding from MIUR (Italy), PRIN Grant No. 20173B72NB, from the European Union, project MaX Materials design at the eXascale H2020-EINFRA-2015-1, (Grants Agreement No. 824143), and project Nanoscience Foundries and Fine Analysis-Europe H2020-INFRAIA-2014-2015 (Grant Agreement No. 654360).

\bibliography{biblio}

\end{document}